\documentstyle[aps]{revtex}
\begin{document}
\draft
\title{Nonlinear coherent wavetrains and bifurcations in a class of deformable models for one-dimensional nonlinear discrete Klein-Gordon systems}
\author{Alain M. Dikand\'e} 
\address{Centre de Recherches sur les Propri\'et\'es \'Electroniques des Mat\'eriaux Avanc\'es, Facult\'e des Sciences,
Universit\'e de Sherbrooke J1K-2R1 Sherbrooke-Quebec, CANADA \\ electronic mail: amdikand@physique.usherb.ca}
\date{\today}
\maketitle
\begin{abstract}
The discrete static properties of a class of deformable double-well
 potential models are investigated. The Peierls-stress potential of the models is explicitely given. 
Numerical analysis of the equation of motion reveal different soliton wavetrain profiles, most of
which are periodic. Soliton wavetrains are also found analytically in terms of 
continuum nonlinear periodic wavefunction solutions then called periodons. 
The periodon stability in a lattice phonon bath is discussed. Looking 
at bifurcation phenomena and routes to chaos for a representative model of 
this class, the return map is derived in terms of a two-dimensional, 
two-parameter map. This map appears to be area preserving, possesses three 
characteristic fixed points with one elliptic, and displays complex bifurcation 
diagrams with hopf-like singularities. The routes to chaos also show complexities
 all due to the interplay of two control parameters.
\end{abstract}
\section{Introduction}
\label{sec:level1}
The soliton concept \cite{1,2,3} is now well understood. However, in the past the subject casted
great controversal, but also rich debates around two main themes: the soliton proper existence as real 
physical object and later, its stability namely against collisions with other objects of 
the same kinds(solitons, phonons) and with hosts(isotopic impurities, etc.). The concensus
 which emerges from these past activities involves the recognition of the universal role that 
soliton and in general, nonlinear waves play in various applications. In condensed matter, 
the soliton concept attracted attention for its intrinsic interest in the fundamental problem of 
phase transitions \cite{1}. In short, the main feature here is that two or more phases of materials may 
coexist separated by well-defined phase boundaries. As phase transition processes are dynamical
 by nature, the great theoretical challenge has been to find appropriate scenarios. The most
 popular one consists to manage at least one of the possible phases and next to follow its evolution
 among different other phases. This scheme takes into account the nucleation, propagation and 
anihilation of phase boundaries. During nucleation, the transition involves by turns an homogeneous, 
single-phase configuration and an inhomogeneous, multiphase configuration. The anihilation takes 
place in the opposite direction. Propagation involves motion of an already existing phase interface, 
which moves through the system increasing or decreasing volume of one phase at the expense of the other. \\
While this scenario applies in various theoretical contexts, the physical systems called in question are
not always the same. For this reason, several one-dimensional(1D)models are used provided acceptable basic structural considerations.
In the theory, each model bears on a specific one-site potential expected to provide the appropriate
features for a given physical system. The one-site potentials which have been widely studied are the $\phi^4$ \cite{5,6},
and sine-Gordon(sG) \cite{7,8}. Both admit nonlinear solutions which are solitons. By its shape profile, the
kink soliton has quite often been associated with phase boundary phenomena \cite{4,9,10}. \\ In practices, phase transitions are
actually more complex and phase boundaries can largely depart from the simplest "single-kink" view-point \cite{11}. To this subject,
relevant insights have been gained in some interesting results namely of Aubry\cite{12}, Bak et al. \cite{13} and
Pokrovsky \cite{14}, from which we learn that discreteness of the propagation medium is an intrinsic deal of the problem.
Thus, lattice discreteness can by themselves determine the becomings of phase boundaries. This leads to relevant phenomena as
"pinned interfaces" and "domain" configurations, revealed by experiments. In addition,
the transient phases can display unexpected behaviours then indicating (phase and/or shape)memory-loss processes. These unpredictible
behaviours are referred to as chaotic regimes, also pointed out and discussed at length \cite{15,16,17,18}. \\
 The recent advance on the soliton theory was directed toward new models with parametrized one-site potentials \cite{19,20,21,22,23,24}. Their advantageous features
can be summarized in the fact that they reproduce the $\phi^4$ and sG while avoiding must of their shortcomings. This paper deals with a class
of deformable double-well potentials(DDWP) introduced very recently \cite{20,22,23}. None of the versions of this class is equivalent to
the familiar sinh-Gordon(shG) potential \cite{25}, instead they must be regarded as distinct generalized versions of the shG model. These DDWP
have several virtues including analytical tractability, in contrast with many others for which it is not always easy to get deeper in
their explicit treatments. Moreover, their "double-well" shapes can be manipulated at one wish and hence allow fitting results to a desired context. To
set these interesting features in a larger perspective, in the present work we will address to some questions related to lattice discreteness. \\
The outlines of the work are the followings: In section II, we review their continuum dynamics and derive explicitely the Peierls stress potential
(so-called Peierls-Nabarro, PN potential) by the Fourier-transform method \cite{12,13,14,15,16,17,18,19,20,21,22,23,24,26}. In section III We examine some configurations of the nonlinear
solutions given by the numerical analyzis. An approximate analytical solution is looked for in terms of periodons, i.e. periodic soliton wavetrains.
The periodon stability in an intrinsic phonon bath is discussed. In section IV the return map of one representative model is derived in terms of
a two-dimensional, two-parameter(2D2P) map evolving in the phase space through distinct bifurcation scenarios in which alternate periodic and chaotic phases. Section V present concluding remarks.

\section{The models, kinks and Peierls-stress potential}
 Consider a one-dimensional($1D$) Klein-Gordon(KG) system, that consists of a chain of harmonically coupled particles with uniform
mass M, each in the well bottom of a doubly-degenerate minimum potential. The dynamics
of the model is described by an Hamiltonian:
\begin{equation}
H= \sum_{n=1} \left[{\frac{1}{2}} \phi^2_{n,t} + {\frac{C^2_o}{2\l^2}} \left(\phi_{n+1}-\phi_n \right)^2 + \omega^2_o V(\phi_{n},\mu)\right] \label{a1}
\end{equation}
Where $\phi_n$ is the particle local displacement field, $C_o$ and $\omega_o$ the bare velocity and frequency respectively, and $V(\phi,\mu)$
 the DDWP that we assume of general form \cite{23}:
 \begin{equation}
V(\phi,\mu)= {\frac{a}{8}} {\left[{\frac{1}{\mu^2}}Sinh^{2}(\alpha \phi)-1\right]}^2 , \hspace{.2in} \mu\neq 0, \hspace{.2in}  a= {\frac{a_{o} q^{2}}{[Sinh^{-1}(\mu)]^2}} \label{a2}
 \end{equation}
The quantity $a_o$ will be constant while $\mu$ can vary. q and $\alpha$ are depending on the main deformability parameter $\mu$ and can
take three distinct values i.e.: \\
\begin{enumerate}
\item[{\it i}])
\begin{equation}
\alpha = \mu,  \hspace{.2in}  q= Sinh^{-1}(\mu) \label{a3}
\end{equation}
For these values, $V(\phi,\mu)$ is a DDWP whose degenerate minima are continuously shifted but leaving unaffected the barrier height \cite{20}. \\
\item[{\it ii}])
\begin{equation}
\alpha = Sinh^{-1}(\mu),   \hspace{.2in} q= {\frac{\mu}{\sqrt{1+\mu{^2}}}} \label{a4}
\end{equation}
For these second values, only the potential barrier will be affected while positions of the degenarate minima remain bounded at $\phi^{o}_{1,2} =\pm{1}$ \cite{22}. \\
\item[{\it iii}])
\begin{equation}
\alpha ={\frac{Sinh^{-1}(\mu)}{\sqrt{1+\mu^2}}},  \hspace{.2in}   q= 1 \label{a5}
\end{equation}
\end{enumerate}
For these last values, the whole "double-well" shape of the potential changes by varying $\mu$ \cite{23}. \\
 In the continuum limit, the generalized DDWP\ (\ref{a2}) can be treated analytically and yields the following single-kink solution:
\begin{equation}
\phi(s=x-vt, \mu)= \pm{\frac{1}{\mu}}\tanh^{-1}{\left[{\frac{\mu}{\sqrt{1+\mu^2}}}tanh{\left({\frac{\gamma s}{{\sqrt{2}}d(\mu)}}\right)}\right]} \label{a6}
\end{equation}
\begin{equation}
 d^{2}(\mu)= {\frac{(1+\mu^{2})d^2_o}{a\alpha^2 \mu^2}}, \hspace{.2in} d^2_o = {\frac{C^2_o}{\omega^2_o}}, \hspace{.2in} \gamma^{-2}= 1-{\frac{v^2}{C^2_o}} \label{a7}
\end{equation}
The associate mass as well, in virtue of its particle-like properties, is analytically obtainable and reads:
 \begin{equation}
M_{k}(\mu)= {\frac{\sqrt{a}}{2\mu^2\alpha d_o}}{\left[Sinh(2\alpha)-(1+2\mu^2)\alpha \right]}   \label{a8}
\end{equation}
This analytical expression of the generalized kink topological mass deserves instructive remarks concerning its defferent behaviours with repect
to $\mu$, for the three cases listed in ${\it i})-{\it iii})$:
 in the first, the kink mass displays a drastic decrease when increasing $\mu$, but remains always positive and non-zero. Similar behaviour
prevails in the second case. In the third, by contrary, the kink mass behaves oppositely, increasing from negative to positive values through
 massless-kink shape for $\mu$ at about 0.76 \cite{23}. \\
 We now turn our attention to the response of the single kink obtained above to lattice discreteness effetcs. We can restrict ourselves to the
 static-kink regime, which is suitable for a "pinned-kink" configuration where kink pinnings are tough to result from forces provided
by the discreteness of the propagation medium. These forces are often identified with the Peierls stresses governing structural changes in solid-state
systems \cite{12,13,14,26}. The generating potential is so-called Peierls-stress potential and has been the subject of a great recent
interest \cite{27}, most often referred to as "Peierls-Nabarro"(PN) pinning potential. \\
In the static regime, the total discrete system energy reduces to:
\begin{equation}
U= \sum_{n=1}{G(n)}, \hspace{.2in} G(n)= {\frac{MC^2_o}{\l^2}}{\left({\frac{d\phi_n}{dn}}\right)^{2}} \label{a9}
\end{equation}
and the PN potential then follows by Fourier transforming G(n), i.e. calculating:
\begin{equation}
U_{PN} = MC^2_{o}\sum^{\infty}_{m=-\infty}{\int^{\infty}_{0}{dn{\left({\frac{d\phi_n}{dn}}\right)^2}} \exp{(2\pi imn)}} \label{a10}
\end{equation}
with:
\begin{equation}
{\left({\frac{d\phi_n}{dn}}\right)^2} = \left({\frac{B^{2}_o}{cosh^2({\frac{n\l}{\sqrt{2}d(\mu)}})+\mu^2}}\right)^2,  \hspace{.2in} B^2_{o} = {\frac{\mu^2(1+\mu^2)}{2d^2(\mu)}} \label{a11}
\end{equation}
It is easy to check that $G(n)$ is an even function of the variable n. Therefore the Fourier harmonics follow from the class of integrals:
\begin{equation}
{\int^{\infty}_{0}{dX{\frac{cos(XY)}{\left[cosh(X)+cosh(X_o) \right]^2}}}}= {\sqrt{\frac{\pi}{2}}}{\frac{\Gamma(2+iY)\Gamma(2-iY)}{\Gamma(2)sinh^{3/2}(X_o)}}\wp^{-3/2}_{-1/2+iY}[cosh(X_o)] \label{a12}
\end{equation}
\begin{equation}
X= {\frac{2n\l}{\sqrt{2}d(\mu)}}, \hspace{.2in} cosh(X_o) = 1+2\mu^2, \hspace{.2in}  Y= \pi \sqrt{2}md(\mu) \label{a13}
\end{equation}
$\wp$ in this relation is the Associate Legendre function here appropriately defined as:
\begin{eqnarray*}
\wp^{-3/2}_{-1/2+iY}[cosh(X_o)]= -{\sqrt{\frac{1}{2\pi}}}{\frac{1}{sinh^{3/2}(X_o)}}{\int^{\infty}_{X_o}{du[cosh(u)-cosh(X_o)]cosh(\pi Y)cos(Yu)}} +
\end{eqnarray*}
\begin{equation}
{\int^{\infty}_{0}{du[cosh(u)+cosh(X_o)]cosh(\pi Y)cos(Yu)}}  \label{a14}
\end{equation}
While the complex-argument Gamma function $\Gamma$ is such that:
\begin{equation}
\Gamma(2+iY)\Gamma(2-iY)= {\frac{ \mid \Gamma(2) \mid^2 }{{\prod^{\infty}_{k=0}} \left[1+{\frac{Y^2}{(2+k)^2}} \right]}} \label{a15}
\end{equation}
with the product function evaluated by standard formula to give:
\begin{equation}
{\prod^{\infty}_{k=0}}{\left[1+{\frac{4Y^{2}}{(2k+1)^2 \pi^2}}\right]} = cosh(Y) \label{a16}
\end{equation}
From now on, combining \ (\ref{a9})-\ (\ref{a16}),  we find:
\begin{equation}
U_{PN}(n) = U_o + \sum^{\infty}_{m=1}{U_m cos(2\pi mn)} \label{a17}
\end{equation}
\begin{equation}
U_o = { \frac{ \mu^2 (1+\mu^2 )C^2_o \l}{2 \sqrt{2} d(\mu)}}{\left[X_{o} cotanh(X_o ) - 2\right]}, \hspace{.2in}
U_m = { \frac{8B^2_{o} C^2_{o} \l }{sinh^{3/2}(X_o)cosh({\frac{m\pi^2 d( \mu)}{2 \sqrt{2} \l}})}}{ \sqrt{ \frac{\pi}{2}}} \wp^{-3/2}_{-1/2+iY}[cosh(X_o)] \label{a18}
\end{equation}
Comparing the zero-harmonic $U_o$ and the kink rest mass obtained previously in\ (\ref{a8}), it appears that this term reproduces the kink proper energy.
The next harmonics will then account for the lattice discreteness effects. The leading term $U_1$ is dominant since the cosh argument is already almost insensitive to the second order in m.

\section{Soliton wavetrains and periodons}
 When investigating the PN potential in section II, we assumed a continuum kink wave solution owing to the difficulty of proceeding with
 analytical treatment of the discrete motion equation. The best way of accessing discrete solutions is by numerical analysis. For
this purpose, we will use the following nonlinear difference equation:
\begin{equation}
\phi_{n+1} + \phi_{n-1} -2\phi_n = {\frac{1}{d^2_o}}{\frac{\partial{V(u_{n}, \mu)}}{\partial{u_n}}} \label{a19}
\end{equation}
where $d_o$ now sets the kink(or in general, soliton)length scale. We will only show the results for one representative among the
DDWP above, that is \ (\ref{a3}). On figures 1 and figures 2 we displays some characteristic features
emerging from the iterates in the phase space and of the wave shapes, respectively. It is manifest that the "wavetrain" feature prevails
in the almost whole discrete regime. In particular, the discrete nonlinear regimes(Figs.2a, 2b) appear to be essentially dominated by
"soliton-lattice" structures. These soliton-lattices solutions describe multi-kink states spreading along the chain axis with well defined
periodicities. By varying the deformability parameter $\mu$, the "kink-lattice" structure collapses gradually and turns into almost harmonic
periodic wavetrains via periodic structures that we can identify as weakly nonlinear,
discrete modulational wavetrains(Fig.2c). \\
Tough characteristic of the discrete nonlinear equation, the kink-lattice structure is not specific to the discrete phase. Indeed, solving the continuum
limit of the nonlinear difference equation provided appropriate boundary conditions such as periodic boundary conditions, we may arrive at a continuum soliton
wavetrain solution then called periodon. In principle, there are two different ways to check that the periodon describes lattice of kinks with well defined periodicity.
The first is by direct integration of the motion equation \cite{28,29,30,31}, leading to Elliptic functions which are indeed periodic. The second
is by a naive and artificial method that consists in an algebraic sum of several(infinite) one-single kink wavefunctions\cite{32}. Both ways generate the same wavetrain profile and hence are equally valid.
However, we choose to proceed via the first and find for the generalized version\ (\ref{a5}):
\begin{equation}
\phi(s=x-vt, \mu)= \pm{\frac{1}{\alpha}}\tanh^{-1}{\left[{\beta_2}Sn{\left({\frac{\gamma s}{{\sqrt{2}}d_{L}(\mu, \lambda)}}\right)}\right]}, \hspace{.2in} 0\leq \lambda \leq 1 \label{a20}
\end{equation}
\begin{eqnarray*}
d(\mu,\lambda)= {\frac{d_o}{\beta_{1} \alpha \sqrt{C}}}, \hspace{.2in}  {\frac{\beta_{1}^2}{\beta_{2}^2}}= \lambda, \hspace{.2in} \beta_{1}^{2} \beta_{2}^{2}= {\frac{A}{C}}, \hspace{.2in} \beta_{1}^2 + \beta_{2}^2 = {\frac{2B}{C}}
\end{eqnarray*}
\begin{equation}
A= {\frac{a}{8}} + K_o , \hspace{.2in} B= {\frac{a(1+\mu^{2})}{8\mu^{2}}} + K_o , \hspace{.2in}  C= {\frac{a(1+\mu^{2})^{2}}{8\mu^{4}}} + K_o \label{a21}
\end{equation}
Sn is the Jacobi Elliptic function of modulus $\lambda$, $K(\lambda)$ the Jacobi Elliptic Integral of the first kind and $K_o$ is self-consistently given by the second-order polynomial:
\begin{equation}
{\frac{a^{2}(1+\mu^{2})^{2}(1-\lambda)^{2}}{64\mu^{4}}} + {\frac{a}{8}} \left[ {\frac{8(1+\mu^{2})\lambda}{\mu^{2}}}- \left(1+{\frac{(1+\mu^{2})^{2}}{\mu^{4}}} \right)(1+\lambda)^{2} \right] K_{o} + (1-\lambda)^{2}K_{o}^{2} =0 \label{a23}
\end{equation}
The period of the nonlinear solution \ (\ref{a20}) is:
\begin{equation}
L= d_{L}(\mu, \lambda)K(\lambda) \label{a24}
\end{equation}
To clearly see some interesting asymptotic behaviours of \ (\ref{a20}), we consider the two limits $ \lambda \rightarrow 0$ and $ \lambda \rightarrow 1$. The first leads to:
\begin{equation}
\phi(s=x-vt, \mu)= \pm{\frac{1}{\alpha}}\tanh^{-1}{\left[\beta_{2} \sin{\left({\frac{\gamma s}{{\sqrt{2}}d_{L}(\mu, 0)}}\right)}\right]} \label{a25}
\end{equation}
This solution describes a breathing-like excitations, i.e. a kink with internal periodic quasi-linear oscillations. The effect of $\mu$ on the shape of this excitation mode
 is explicit from an expansion of\ (\ref{a25}) about small values of this paramter. The leading term yields:
\begin{equation}
\phi(s=x-vt, \mu)= \pm{\frac{\beta_2}{\alpha}}\sin{\left({\frac{\gamma s}{{\sqrt{2}}d_{L}(\mu, 0)}}\right)} \label{a26}
\end{equation}
For the second limit, expansions of the periodon wavefunction for small values of $ \mu $ give back the single-kink solution\ (\ref{a6}). \\ We would like to end this section with a short
discussion on the stability of the periodon kink above against collisions in a phonon bath. Let us recast the periodon in the form:
\begin{equation}
\phi(s=x-vt, \mu)= \pm{\frac{1}{\alpha}}\tanh^{-1}[u(s)], \hspace{.2in} u(s)= \beta_{2}Sn{\left({\frac{\gamma s}{{\sqrt{2}}d_{L}(\mu, \lambda)}}\right)}, \hspace{.2in} 0\leq \lambda \leq 1 \label{a27}
\end{equation}
 where the function $u(s)$ is the part of the periodon wavefunction undergoing "shape dressing" upon collisions in the phonon bath. Thus we have to introduce the periodon-phonon ansatz
as follows:
\begin{equation}
u(s)= u(x)+ \psi(x) exp(\it i \omega t) \label{a28}
\end{equation}
The resulting eigenvalue problem is then:
\begin{equation}
\psi_{zz}+ {\left[P(\Omega,\lambda)- 6 \lambda Sn^{2}(z) \right]}\psi=0, \hspace{.2in} z= {\frac{x}{\sqrt{2}d_{L}(\mu)}}, \hspace{.2in} \Omega^{2}= {\left({\frac{\omega}{\omega_{o}}} \right)^{2}}{\frac{1}{4B\alpha^{2}}},
\hspace{.2in} P(\Omega,\lambda)= (1+\Omega^{2})(1+\lambda) \label{a29}
\end{equation}
This is just the Jacobian form of the (second-order) Lam\'e equation\cite{33,34}. The same equation is found for the $\phi^4$ in the finite-support(or finit period) limit\cite{33}. Its bound state
spectrum possesses four non-zero frequency modes in addition to the translational mode.

\section{Bifurcations and routes to chaos}
Modulated systems appear as the most frequent candidates when evoking systems that can be modelled within the present theory. These systems have long served as
 prototypes of materials with complex behaviours. The modulation periods are either commensurate or incommensurate with the basic lattice then providing them
with their unique and rich physical properties. The transition between the two phases involve domain walls which are pinned soliton lattices. Properties of these
soliton lattices are fundamental in determining the nature of the prevailing phase. It has been shown \cite{11,12,13,14,15,16,17,18} that the appearance of the soliton lattice can occur simultaneously
with new phases which are essentially metastable or strongly unstable. Among them, lock-in and chaotic regimes correspond respectively to periodic domain-wall lattice and randomly pinned
domain-wall lattice configurations. Theoretical attempts
 to understand these chaotic regimes have been carried for both $\phi^4$ and sG, in connection with physical phenonema as Peierls instabilities. \\
To a general viewpoint, the possibility of describing classical mechanical systems by discrete difference equations provides an excellent framework where
to connect the well-understood periodic and chaotic behaviours of deterministic models with the equivalent behaviours in discrete nonlinear KG systems. The discrete logistic
map is a sound ground for
this purpose. According to existing approaches, the system will be said to exhibit chaos when its phase space trajectories show initially exponential
sensitivity to small differences in initial conditions. This feature can be made quantitative by showing that the largest Lyapunov exponent is positive.
We already know that for classical Hamiltonian
dynamics with sympletic structures, the sum of all of the Lyapunov exponents is zero such that at least one is positive, unless all vanish as they due for
integrable systems. By these means, chaotic instabilities are virtually generic for non integrable conservative Hamiltonian systems. In a past work \cite{35},
 Bak and Jensen have considered the problem for the $\phi^4$,
showing that the theory could be formulated as an area-preserving $2D$ mapping. Their approaches cover a wide range of models involving those dealth with in the present context. Still, the
competitions between the two main parameters(a and $\mu$)
in the DDWP are likely to generate new singularities in both bifurcation diagrams and chaotic windows and it is the aim of this section to point out some of the revelant ones.    \\
Starting, consider the $2D2P$ map:
\begin{equation}
M_{(a,\mu)}: $$\cases{u_{n+1}= -v_n + 2u_n + {\frac{\partial V(u_{n},\mu)}{\partial u_n}}, \cr
 v_{n+1}= u_{n}\cr}$$ \label{a30}
\end{equation}
$M_{(a,\mu)}$ can evolve among stable and unstable equilibra which are the fixed points of the map, and any set of points to form a trajectory on the
map area will result in a periodic or chaotic orbit of one of these fixed points. As it sounds, whether the fixed points are stable or not is relevant in determining
the local behabiours of the map area.
Thus, a stable fixed point will be such that the orbits near it are moved even closer under the map, whereas nearby orbit points of the unstable fixed
point will move aways by iteration. We can understand these equilibrium behaviours in the simple viewpoint of linear stability analysis, which consists
to look at the transformation under $M_{(a,\mu)}$ as "gradient crossings"
 of neighbouring orbit points on the surface area consisting of the whole trajectory set of the map. Mathematically, we end with the matrix equation:
\begin{equation}
M_{(a,\mu)}: \hspace{.4in} M_{n}W_{n}= W_{n+1} \label{a31}
\end{equation}
Where $M_n$, then called Jordan matrix of the map, is the square Jacobian matrix of the gradients components at points $W_{n}={\left[u_{n}, v_{n} \right]}$ on the map area. From
now on, further iteration will result in conjugating the matrix $M_n$ such that for a given period-$\l$ orbit of some fixed point $W_o$, the generator
 of the period-$\l$ orbit points is $M^{\l}_o$. It turns out that properties of the transformation matrix will have direct incidence on the geometry of the map area.
Indeed, since a matrix with determinant of absolute value
different from one traduces change of norm, for the map this will correspond either to the contraction or to the expansion of the map area about the
fixed point. Else, the map will be invariant under transformation and hence is area preserving.
 On the other hand, if all the eigenvalues of the matrix have magnitudes less(greater) than one, the associate fixed point will always attract(repel)
the orbit points and therefore is and attractor(repeller). If one eigenvalue is less than one and the other is greater, the fixed point is a saddle:
 any perturbation of the orbit away from it may be magnified by further iteration leading to an unstable fixed point.
 The fixed point will be hyperbolic if none of the eigenvalues has magnitude one, and elliptic if the two eigenvalues lie on a unit circle one the map
 area. In theory, bifurcations of any of these fixed(or of any other periodic)points on the map can occur only for the parameter values for which an
 eigenvalue of the corresponding orbit set has absolute value one.
Basing on this preliminary, we start by looking for the fixed points of the map $M_{(a,\mu)}$. They are three:
\begin{equation}
W^{(o)}=(0,0), \hspace{.2in} W^{\it i}= (-1)^{\it i}{\left[{\frac{Sinh^{-1}}{\mu}}, {\frac{Sinh^{-1}}{\mu}} \right]}_{\it{i} =1,2}  \label{a32}
\end{equation}
It is noted that they coincide exactly with the fixed points of the 1D invariant subspace of the map. The Jordan matrix for the period-one orbit set is the $2D$ matrix:
\begin{equation}
M_{n}= \left( \begin{array}{cc}2+ {\frac{a Sinh(2\mu)Sinh(2\mu x)}{\mu^{2}}} & -1 \\ 1 & 0 \end{array} \right) \label{a33}
\end{equation}
The norm of its determinant is one whatever the values of a and $\mu$: therefore the map is always area preserving.
Its eigenvalues at the fixed points $W^{o}, W^{\it i}$ are respecively:
\begin{equation}
W^{o}: \hspace{2in} {\left(e^{o}_{1}, e_{2}^{0}\right)}= (1, 1) \label{a34}
\end{equation}
 \begin{equation}
W^{+}: $$\cases{ e^{+}_{1}= 1+{\frac{a \sqrt{1+\mu^{2}}}{\mu}}Sinh(2\mu)+ \sqrt{{\frac{a \sqrt{1+\mu^{2}}}{\mu}}\left[{\frac{a \sqrt{1+\mu^{2}}}{\mu}}Sinh(2\mu)-2 \right]Sinh(2\mu)}  \cr
 e^{+}_{2}= 1+{\frac{a \sqrt{1+\mu^{2}}}{\mu}}Sinh(2\mu)- \sqrt{{\frac{a \sqrt{1+\mu^{2}}}{\mu}}{\left[{\frac{a \sqrt{1+\mu^{2}}}{\mu}}Sinh(2\mu)-2 \right]}Sinh(2\mu)} \cr}$$ \label{a35}
\end{equation}
\begin{equation}
W^{-}: $$\cases{ e^{-}_{1}= 1-{\frac{a \sqrt{1+\mu^{2}}}{\mu}}Sinh(2\mu)- \sqrt{{\frac{a \sqrt{1+\mu^{2}}}{\mu}} \left[{\frac{a \sqrt{1+\mu^{2}}}{\mu}}Sinh(2\mu)-2 \right]Sinh(2\mu)} \cr
 e^{-}_{2}= 1-{\frac{a \sqrt{1+\mu^{2}}}{\mu}}Sinh(2\mu)+ \sqrt{{\frac{a \sqrt{1+\mu^{2}}}{\mu}}{\left[{\frac{a \sqrt{1+\mu^{2}}}{\mu}}Sinh(2\mu)-2 \right]}Sinh(2\mu)} \cr}$$ \label{a36}
\end{equation}
Where the signs $\mp$ stand for $\it i$=1, 2. While the fixed point $(0, 0)$ is elliptic, for the two others the asymptotic behaviours
 are not so clear. We can circumvent this uncertainty by considering instead the traces of the matrix at those points. We obtain:
\begin{equation}
 T^{+}_{1(a,\mu)} = 2 \left[1+{\frac{a\sqrt{1+\mu^{2}}}{\mu}}Sinh(2\mu) \right], \hspace{.2in}  T^{-}_{1(a,\mu)}= 2 \left[1-{\frac{a\sqrt{1+\mu^{2}}}{\mu}}Sinh(2\mu) \right]  \label{a37}
\end{equation}
In terms of these traces and according to our preliminary statments, the stability condition can now turn into the inequality: $ \mid T_{1(a,\mu)}^{\pm} \mid  <  2$.
Since $\mu$ and a are all positive definite, it results that this condition on the trace will never hold for $T_{1(a,\mu)}^{+}$  and therefore $W^{+}$ is an always unstable fixed point.
By contrary, $\mid T_{1(a,\mu)}^{-} \mid$ can be less than 2 provided we select the following parameter values:
\begin{equation}
a  >  {\frac{{\mu}}{\sqrt{1+\mu^{2}}Sinh(2\mu)}} \label{a38}
\end{equation}
In which case $W^{-}$ is a stable fixed point. Otherwise, it is unstable. The marginal stability at
\begin{equation}
 a  =  {\frac{{\mu}}{\sqrt{1+\mu^{2}}Sinh(2\mu)}}, \label{a39}
\end{equation}
marks a critical state at which the fixed point can bifurcates carrying the orbit either aways from it or
closer by changing period. Since similar features hold for the fixed point $(0, 0)$, everywhere below we will deal with trace. \\
The stability of the next orbits, i.e. the period-two orbit and higher can from now be easily analyzed. Besides we are able to construct the recusion laws governing successive orbits
around each fixed point. Indeed,
writing the matrix equation for the transformation to a given \l -cycle orbit as $W = M^{\l}W$, we find by turns:
\begin{equation}
T^{o}_{\l (a,\mu)}= 2^{\l}, \hspace{.2in}  T^{+}_{\l (a,\mu)}= {\left[2{\left(1+{\frac{a\sqrt{1+\mu^{2}}}{\mu}}Sinh(2\mu) \right)} \right]}^{\l}, \hspace{.2in} T^{-}_{\l (a,\mu)}=
{\left[2{\left(1-{\frac{a\sqrt{1+\mu^{2}}}{\mu}}Sinh(2\mu) \right)} \right]}^{\l} \label{a40}
\end{equation}
  We learn from these recursion relations that the sequences of orbit cycles are indeed all periodic and therefore result in the map dynamics characterized
 by changes in the geometry of the KAM surfaces. As the phase-space trajectories in figures 1 illustrate, these per-iterate changes involve stretchings of the two axes as the separations
between nearby points increase or decrease. The stretching factors in each of these two directions can be identified as the largest Lyapounov exponents of the orbit. For
the fixed point $(0,0)$, the recursion suggests that the norms of the largest Lyapunov exponents of successive orbits all converge to the same limit
$L= \ln(2)$. Positive Lyapunov exponent traduces the fact that bifurcations of the fixed points will consist of period doublings(or halvings) as
long as orbits remain asymptotically periodic, and can involve openings of chaotic windows along the bifurcation loop. By the same means we can estimate the
convergence limits of the largest Lyapunov exponents of the fixed points $W^{\pm}$. we find one single expression for the two non-zero fixed points\ (\ref{a35})-\ (\ref{a36}) i.e:
\begin{equation}
L^{(\it i)} = \ln(2)+  \ln{\left[1-{\frac{a^{2}(1+\mu^{2})}{\mu^{2}}}Sinh^{2}(2\mu)\right]} \label{a41}
 \end{equation}
 According to figure 3a and 3b , depending on the values of a and $\mu$ the parameter $L^{(\it i)}$ can change
from positive to negative values and crosses a zero value without discontinuity. This behaviour reminds Hopf singularities. \\ To fix our minds as
concerns the map sensitivity to the two parameters a($a-M_{\mu}$ map) and $\mu$($\mu-M_{a}$ map) taken separately, we follow an
arbitrary trajectory set from an input
position for some range of values of each of these parameters. This leads to the bifurcation diagrams displayed on figure 4 and figures 5a-5c. Figures 4 corresponds
to the $a-M_\mu$ map, with input $u_{o} = 10^{-7}$ and for $\mu = 0.5$. Periods of the orbits seem to get halved converging to the critical
value $a= 1.685$. Next, the first period doubling points out at $ a= 1.963$, followed by a cascade of
further period-doubling bifurcations with alternating chaotic and periodic attractor windows. Namely, the period-six attractor window is
observed at about $a = 2.165$. Similar features also emerge for the $\mu-M_a$ map. However, here emerges two main bifurcation scenarios
shown on figures 5a and 5b, respectively.
Figure 5a corresponds to values of a in the range $a< 1$(in this particular case $a=0.99$). The first crisis consists of period doubling
at $\mu = 0.76$ with the period-six chaotic attractor windows settling at $\mu= 0.937$. For $a>1$, the prevailing bifurcation scenario is that
shown on figure 5b($a=1.99$). On figure 5c we magnify the region of figure 5b that shows apparent crisis.
Here also we note "period-doubling" cascades as well as chaotic phases marked by the random points.
These complex behaviours suggest that the $\mu-M_a$ map may show deeper sentitivity to parameter changes than the $a-M_\mu$.
\section{Conclusion}
\label{sec:Level5}
In summary, we have explored the essential discrete properties of a class of DDWP models. We have been able
to derive analytical expressions of some important physical quantities as the PN potential, which makes the DDWP analytically
tractable an therefore applicable to a large variety of discrete nonlinear systems. The multiplicity and richness of the distinct discrete wave
amplitude solutions resulting from the numerical simulations, as well as the bifurcation diagrams showing
the coexistence of period-doubling and period-halving cascades as well as chaotic phases, are of great
importance for the possible large variety of domain-wall and phase regimes that they allow to explore within one single model.
\acknowledgments
Work jointly supported by FCAR-CANADA and ICTP-IAEA-UNESCO, Trieste-Italy.

\newpage

\section{Figure Captions:}
Figure 1a: Phase-space trajectories for $\mu \sim 1.5$, $a \sim 0.8$. \\
Figure 1b: Phase-space trajectories for $\mu \sim 0.1$, $a \sim 0.8$. \\
Figure 2a: $\mu \sim 2$ . Well-shaped discrete periodic soliton wavetrains given by numerical simulations. \\
Figure 2b: $\mu \sim 0.5$ . Discrete periodic wavetrain of narrow solitons. \\
Figure 2c: $\mu \sim 0.01$. Weakly nonlinear, discrete periodic envelope modes. \\
Figure 3a: Largest Lyapunov exponent of the $W^{\pm}$ fixed points as function of $\mu$. \\
Figure 3b: Largest Lyapunov exponent of the $W^{\pm}$ fixed points as function of a. \\
Figure 4: Bifurcation diagram of the $a-M_{(a,\mu)} $ map. \\
Figure 5a: Bifurcation diagram of the $\mu-M_{(a,\mu)}$ map for $a < 1$. \\
Figure 5b: Bifurcation diagram of the $\mu-M_{(a,\mu)}$ map for $a > 1$. \\
Figure 5c: First critical region of the Bifurcation diagram of $\mu-M_{(a,\mu)}$. 
\end{document}